\documentstyle[10pt,psfig,float]{mn}

\title[Nonlinear hydrodynamical evolution of rotating relativistic stars]
      {Nonlinear hydrodynamical evolution of rotating relativistic stars: 
       Numerical methods and code tests}
\author[Jos{\'e} A. Font, Nikolaos Stergioulas and Kostas D. Kokkotas]
       {Jos{\'e} A. Font$^1$, Nikolaos Stergioulas$^{1,2}$ and Kostas D. 
        Kokkotas$^2$ \\
        $^1$Max-Planck-Institut f{\"u}r Gravitationsphysik \\
        Albert-Einstein-Institut \\
        Am M{\"u}hlenberg 5, D-14476, Golm, Germany \\
        $^2$Department of Physics \\
        Aristotle University of Thessaloniki \\
        Thessaloniki 54006, Greece}

\date{}
\pagerange{}
\pubyear{}

\begin{document}

\maketitle

\label{firstpage}

\begin{abstract}
  We present numerical hydrodynamical evolutions of rapidly rotating
  relativistic stars, using an axisymmetric, nonlinear relativistic
  hydrodynamics code. We use four different high-resolution
  shock-capturing (HRSC) finite-difference schemes (based on
  approximate Riemann solvers) and compare their accuracy in
  preserving uniformly rotating stationary initial configurations in long-term
  evolutions. Among these four schemes, we find that the third-order
  PPM scheme is superior in maintaining the initial rotation law in
  long-term evolutions, especially near the surface of the star. It is
  further shown that HRSC schemes are suitable for the evolution of
  perturbed neutron stars and for the accurate identification (via
  Fourier transforms) of normal modes of oscillation. This is
  demonstrated for radial and quadrupolar pulsations in the
  nonrotating limit, where we find good agreement with frequencies
  obtained with a linear perturbation code. The code can be used for 
  studying small-amplitude or nonlinear pulsations of differentially 
  rotating neutron stars, while our present results serve as testbed 
  computations for three-dimensional general-relativistic evolution codes.
\end{abstract}


%%%%%%%%%%%%%%%%%%%%%%%%%%%%%%%%%%%%%%%%%%%%%%%%%%%%%%%%%%%%%%%%%%%%%%%%%
%                                                                       %
\section{Introduction}                                                  %
\label{intro}                                                           %  
%                                                                       %
%%%%%%%%%%%%%%%%%%%%%%%%%%%%%%%%%%%%%%%%%%%%%%%%%%%%%%%%%%%%%%%%%%%%%%%%%

The hydrodynamical evolution of neutron stars in numerical relativity has 
been the focus of many research groups in recent years (see e.g.  Font et al.
1999; Miller, Suen \& Tobias 1999; Mathews, Marronetti \& Wilson 1998;
Shibata 1999a, Nakamura \& Oohara 1998; Baumgarte, Hughes \& Shapiro
1999; Neilsen \& Choptuik 1999). So far, these studies have been
limited to initially non-rotating stars\footnote{After this work was
  completed, we received a preprint by Shibata 1999b, in which
  evolutions of approximate solutions of rotating stars are shown.}.
However, the numerical investigation of many interesting astrophysical
applications, such as the rotational evolution of proto-neutron stars
and merged neutron stars or the simulation of gravitational radiation
from unstable pulsation modes, requires the ability of accurate
long-term evolutions of rapidly rotating stars. We thus present here
an extensive investigation of the nonlinear hydrodynamical evolution
of rotating neutron stars (first results of this study were presented
in Stergioulas, Font \& Kokkotas 1999). As we focus on the
implementation and problems of the hydrodynamical evolution of
initially stationary configurations in stable equilibrium, we do not
evolve the spacetime. This approximation allows us to evolve
relativistic matter for a much longer time than present coupled
spacetime plus hydrodynamical evolution codes. When studying the
evolution of perturbed rotating neutron stars, the approximation of a
static spacetime still allows for qualitative conclusions to be drawn,
since pulsations of neutron stars are mainly a hydrodynamical process.

The rotational evolution of neutron stars can be affected by several
instabilities (see Stergioulas 1998 for a recent review). If hot
proto-neutron stars are rapidly rotating, they can undergo a dynamical
bar-mode instability (Houser, Centrella \& Smith 1994).  When the
neutron star has cooled to about $10^{10}$K after its formation, it
can be subject to the Chandrasekhar-Friedman-Schutz instability
(Chandrasekhar 1970; Friedman \& Schutz 1978) and it becomes an
important source of gravitational waves. It was recently found that
the $l=m=2$ $r$-mode has the shortest growth time of the instability
(Andersson 1998; Friedman \& Morsink 1998) and it can transform a
rapidly rotating newly-born neutron star to a Crab-like
slowly-rotating pulsar within about a year after its formation
(Andersson, Kokkotas \& Schutz 1999; Lindblom, Owen \& Morsink 1998).
In this model, there are several important questions still to be
answered: What is the maximum amplitude that an
unstable $r$-mode can reach (limited by nonlinear saturation)? Is
there significant transfer of energy to other stable or unstable modes
via nonlinear couplings? (Owen et al. 1998).
Does the $r$-mode evolution in a uniformly rotating star lead to differential
rotation? (Spruit 1999, Rezzolla et al. 1999, Friedman 1999).
Such questions cannot be answered by
computations of normal modes of the linearized pulsation equations,
but require nonlinear effects to be taken into account. In addition,
the computation of the frequency of $r$-modes in a rapidly rotating
relativistc star may be easier to achieve in nonlinear numerical
evolutions than in a perturbative computation. It is therefore
desirable to develop the capability of full nonlinear numerical
evolutions of rotating stars in relativity. The present
implementation of the hydrodynamical evolution is a first step in this
direction.

For our study we have developed an axisymmetric numerical code 
in spherical polar coordinates which uses
high-resolution shock-capturing (HRSC) finite-difference schemes for
the numerical integration of the general relativistic hydrodynamic
equations (see Ib{\'a}{\~n}ez \& Mart\'{\i} 1999 for a recent review of
applications of HRSC schemes in relativistic hydrodynamics).  Such
schemes have been succesfully used before, in the study of the
numerical evolution and gravitational collapse of nonrotating neutron
stars in 1-D (Romero et al. 1996).  An alternative approach, based on
pseudospectral methods, has been presented by Gourgoulhon (1991).
Using our code in 1-D, we can accurately identify radial normal modes
of pulsation up to high harmonic order.  In 2-D the code is
suitable for the hydrodynamical evolution of rotating stars and the
study of axisymmetric modes of pulsation.

We find that the evolution of rotating stars with HRSC schemes can be
subject to a significant numerical difficulty at the surface of the
star. This difficulty arises due to the fact that one of the evolved
hydrodynamical variables, which corresponds to the angular momentum, 
has a sharp discontinuity at the surface. The result of a
time-dependent evolution is a secular drift of the
evolved configuration from the initial rotation law, near the surface.
We investigate this problem, using two different classes of HRSC
schemes: Total Variation Diminishing (TVD, Harten 1984) and
Essentially Non-Oscillatory (ENO, Harten \& Osher 1987; Harten et al.
1987). We use both second and third-order variations of these schemes
and find that, for a given resolution, the third-order schemes are
superior in dealing with this problem. In addition, we show that all
schemes are suitable for identiying normal modes of pulsation in a
hydrodynamical evolution, via Fourier transforms of the evolved
variables.

In a subsequent paper we will present frequencies of quasi-radial and
axisymmetric pulsations of rapidly rotating neutron stars. We further
plan to study the pulsations of differentially rotating, hot
proto-neutron stars and determine possible nonlinear couplings of
different modes of pulsation.

The paper is organized as follows: In Section 2 we describe the
initial equilibrium configurations which will be evolved with the
hydrodynamical code. In Section 3, we present the explicit form of the
equations of general relativistic hydrodynamics as implemented in our
code. Section 4 is devoted to the description of the numerical
implemetation of the evolution schemes. 
Finally, Section 5 presents 1-D and 2-D tests of the code with the various
implemented schemes.

%%%%%%%%%%%%%%%%%%%%%%%%%%%%%%%%%%%%%%%%%%%%%%%%%%%%%%%%%%%%%%%%%%%%%%%%%
%                                                                       %
\section{Initial Configurations}                                        %
\label{initial}                                                         %  
%                                                                       %
%%%%%%%%%%%%%%%%%%%%%%%%%%%%%%%%%%%%%%%%%%%%%%%%%%%%%%%%%%%%%%%%%%%%%%%%%

Our initial models are exact numerical solutions of rapidly rotating
relativistic stars, having uniform angular velocity $\Omega$. The
metric of the stationary and axisymmetric spacetime in quasi-isotropic 
coordinates is
\begin{eqnarray}
ds^2 &=& -e^{2 \nu}dt^2 + B^2e^{-2\nu}r^2\sin^2\theta(d\phi-\omega dt)^2
\nonumber \\ &&
+e^{2\alpha}(dr^2+r^2d\theta^2),
\label{metric}
\end{eqnarray}
where $\nu$, $B$, $\alpha$ and $\omega$ are metric functions (Butterworth \&
Ipser 1976). In the non-rotating limit the above metric reduces to
the metric of spherical relativistic stars in isotropic coordinates.
We use dimensionless quantities by setting $c=G=M_\odot=1$.

We assume a perfect fluid, zero-temperature equation of state (EOS),
for which the energy density is a function of pressure only. The
following relativistic generalization of the Newtonian polytropic EOS
is chosen:
\begin{eqnarray}
p&=&K\rho_0^{1+1/N}
\\
\epsilon&=& \rho_0 +Np,
\end{eqnarray}
\noindent
where $p$ is the pressure, $\epsilon$ is the energy density, $\rho_0$ is the
rest-mass density, $K$ is the polytropic constant and $N$ is the
polytropic exponent. This form of a relativistic polytropic EOS is
convenient to use, since it coincides with the ideal fluid EOS used
in many relativistic hydrodynamical codes, in the case of an adiabatic
evolution.

The initial equilibrium models are computed using a numerical code by
Stergioulas \& Friedman (1995) which follows the Komatsu, Eriguchi \&
Hatchisu (KEH, 1989) method (as modified in Cook, Shapiro \&
Teukolsky, CST, 1992) with some changes for improved accuracy.  In the
KEH method three of the four general-relativistic,
partial-differential field equations are converted to integral
equations using appropriate Green's functions. The boundary conditions
at infinity are thus incorporated in the integral evaluation. We use a
compactified coordinate (introduced by CST) that allows one to
integrate over the whole spacetime. The code has been shown to be
highly accurate in extensive comparisons to other numerical codes
(Nozawa et al. 1998). A public domain version is available at the
following URL address:
$\verb+http://www.gravity.phys.uwm.edu/Code/rns+$.

The initial model is supplemented by a uniform, nonrotating
``atmosphere'' of very low density (typically $10^{-6}$ or less times
the central density of the star). This is necessary for computing
non-singular solutions of the hydrodynamic equations everywhere in the
computational domain.  After each time-step, we reset the density,
pressure and velocity in the atmosphere to their initial values.  The
influence of the atmosphere is thus minimized and restricted to
the grid-cells through which the surface of the star passes. The
radial velocity of the atmosphere is only set to zero after each
time-step, if it is negative. This avoids unwanted accretion of the 
material in the atmosphere onto the star and at the same time allows 
the star to expand, during radial pulsations.

%%%%%%%%%%%%%%%%%%%%%%%%%%%%%%%%%%%%%%%%%%%%%%%%%%%%%%%%%%%%%%%%%%%%%%%%%
%                                                                       %
\section{General Relativistic Hydrodynamic Equations}                   % 
\label{hydro}                                                           % 
%                                                                       %
%%%%%%%%%%%%%%%%%%%%%%%%%%%%%%%%%%%%%%%%%%%%%%%%%%%%%%%%%%%%%%%%%%%%%%%%%

The equations of general relativistic hydrodynamics are obtained from
the local conservation laws of density current $J^{\mu}$ and
stress-energy $T^{\mu\nu}$
\begin{eqnarray}
\nabla_{\mu}J^{\mu}&=&0, \\
\nabla_{\mu}T^{\mu \nu}&=&0
\end{eqnarray}
\noindent
with
\begin{eqnarray}
J^{\mu}&=&\rho_0u^{\mu},
\\
T^{\mu \nu}&=&\rho_0hu^{\mu}u^{\nu} + p g^{\mu \nu},
\end{eqnarray}
\noindent
for a general EOS of the form $p=p(\rho,\varepsilon)$. Greek (Latin) indices run
from 0 to 3 (1 to 3). This choice of the stress-energy tensor limits
our study to perfect fluids.

In the previous expressions $\nabla_{\mu}$ is the covariant derivative
associated with the four-dimensional metric $g_{\mu\nu}$,
$u^{\mu}$ is the fluid 4-velocity and $h$ is the specific enthalpy
\begin{eqnarray}
h=1+\varepsilon+\frac{p}{\rho_0}
\end{eqnarray}
\noindent
with $\varepsilon$ being the specific internal energy, related to
the energy density $\epsilon$ by
\begin{eqnarray}
\varepsilon=\frac{\epsilon}{\rho_0}-1.
\end{eqnarray}
\noindent

With an appropriate choice of matter fields the equations of
relativistic hydrodynamics constitute a hyperbolic system and can be
written in a flux conservative form, as was first shown by Mart\'{\i},
Ib{\'a}{\~n}ez and Miralles (1991) for the one-dimensional case.  The
knowledge of the characteristic fields of the system allows the
numerical integration to be performed by means of advanced HRSC
schemes, using approximate Riemann solvers (Godunov-type methods).
The multidimensional case was studied by Banyuls et al. (1997) within
the framework of the 3+1 formulation.  Further extensions of this work
to account for {\it dynamical} spacetimes, described by the full set
of Einstein's non-vacuum equations, can be found in Font et al. (1999).
Fully {\it covariant} formulations of the hydrodynamic equations
(i.e., not restricted to {\it spacelike} approaches) and also adapted
to Godunov-type methods, have been recently presented by Papadopoulos 
and Font (1999).

In the present work we use the hydrodynamic equations as formulated in
Banyuls et al. (1997). Specializing for the metric given by
Eq.~(\ref{metric}), the 3+1 quantities read
\begin{eqnarray}
\tilde{\alpha} &=& e^{\nu},
\\
\beta_{\phi} &=& -\omega B^2 e^{-2\nu} r^2 \sin^2\theta,
\\
\gamma_{rr} &=& e^{2\alpha},
\\
\gamma_{\theta\theta} &=& r^2 e^{2\alpha},
\\
\gamma_{\phi\phi} &=& B^2 e^{-2\nu} r^2 \sin^2\theta,
\end{eqnarray}
\noindent
where $\tilde{\alpha}$ is the lapse function (the tilde is
used to avoid confusion with the metric potential $\alpha$),
$\beta_{\phi}$ is the azimuthal shift and $\gamma_{ij}$ is the 3-metric
induced on each spacelike slice.

The (axisymmetric) hydrodynamic equations are written as a first-order
flux conservative system of the form
\begin{eqnarray}
\frac{\partial {\bf u}}{\partial t} +
\frac{\partial \tilde{\alpha} {\bf f}^r} {\partial r} +
\frac{\partial \tilde{\alpha} {\bf f}^{\theta}} {\partial \theta} =
{\bf s},
\label{hydro_system}
\end{eqnarray}
\noindent
expressing the conservation of mass, momentum and energy,
where ${\bf u}, {\bf f}^r, {\bf f}^{\theta}$ and ${\bf s}$ are,
respectively, the state vector of evolved quantities, the radial and polar
fluxes and the source terms. More precisely, they take the form
\begin{eqnarray}
{\bf u} &=& (D,S_r,S_{\theta},S_{\phi},\tau),
\\
{\bf f}^r &=& (Dv^r, S_rv^r+p, S_{\theta}v^r, S_{\phi}v^r, (\tau+p)v^r),
\\
{\bf f}^{\theta} &=& (Dv^{\theta}, S_rv^{\theta}, S_{\theta}v^{\theta}+p,
S_{\phi}v^{\theta}, (\tau+p)v^{\theta}).
\end{eqnarray}
\noindent
The source terms can be decomposed in the following way
\begin{eqnarray}
{\bf s} = \tilde{\alpha} {\bf s}^{\star} - 
\tilde{\alpha} {\bf f}^r \frac{\partial \log\sqrt{\gamma}}{\partial r}
-\tilde{\alpha} {\bf f}^{\theta} \frac{\partial \log\sqrt{\gamma}}
{\partial \theta}
\end{eqnarray}
\noindent
with $\gamma=\det\gamma_{ij}$ and
\begin{eqnarray}
{\bf s}^{\star} &=& \left(0, T^{\mu\nu}\left[\frac{\partial g_{\nu j}}
{\partial x^{\mu}} - \Gamma^{\delta}_{\mu\nu}g_{\delta j}\right],
\right. \nonumber \\ &&
\left.
\tilde{\alpha}\left[T^{\mu t}\frac{\partial\log\tilde{\alpha}}
{\partial x^{\mu}} - T^{\mu\nu}\Gamma^t_{\mu\nu}\right]\right)
\end{eqnarray}
\noindent 
with $j=r,\theta,\phi$. Quantities $\Gamma^{\delta}_{\mu\nu}$ stand for
the four-dimensional Christoffel symbols.  The definitions of the evolved 
quantities in terms
of the ``primitive" variables ${\bf w}=(\rho_0,v_j,\varepsilon)$ are
\begin{eqnarray}
D &=& \rho_0 W,
\\
S_j &=& \rho_0 h W^2 v_j,
\\
\tau &=& \rho_0 h W^2 - p - D,
\end{eqnarray}
\noindent
where $W$ is the relativistic Lorentz factor
\begin{eqnarray}
W \equiv \tilde{\alpha} u^t = \frac{1}{\sqrt{1-v^2}}
\end{eqnarray}
\noindent
with $v^2=\gamma_{ij}v^iv^j$. The 3-velocity components are obtained
from the spatial components of the 4-velocity in the following way
\begin{eqnarray}
v^i=\frac{u^i}{W}+\frac{\beta^i}{\tilde{\alpha}}.
\end{eqnarray}

%%%%%%%%%%%%%%%%%%%%%%%%%%%%%%%%%%%%%%%%%%%%%%%%%%%%%%%%%%%%%%%%%%%%%%%%%
%                                                                       %
\section{Numerical schemes for the hydrodynamic equations}              %   
\label{numerical}                                                       %
%                                                                       %
%%%%%%%%%%%%%%%%%%%%%%%%%%%%%%%%%%%%%%%%%%%%%%%%%%%%%%%%%%%%%%%%%%%%%%%%%

We turn now to the description of the numerical implemention of the
equations presented in the previous section. As we are investigating
the accuracy of the evolution of rotating neutron stars with different
schemes, we include here a detailed technical description of the
different numerical methods and tools we have implemented in our code.
A point we emphasize is the use of high-order polynomial cell-reconstruction 
procedures, to achieve a better representation of the angular-momentum 
discontinuity at the surface of the star.

\subsection{A high-resolution shock-capturing algorithm}
%=======================================================

Our hydrodynamical code performs the numerical integration of system
(\ref{hydro_system}) using Godunov-type (HRSC) methods.  In a HRSC
scheme, the knowledge of the characteristic fields (eigenvalues) of
the equations, together with the corresponding eigenvectors, allows
for accurate integrations, by means of either exact or approximate
Riemann solvers.  These solvers, which constitute the kernel of our
numerical algorithm, compute, at every interface of the numerical
grid, the solution of local Riemann problems (i.e., the simplest
initial value problem with discontinuous initial data). Hence, HRSC
schemes automatically guarantee that physical discontinuities
appearing in the solution, e.g., shock waves, are treated consistently
(the {\it shock-capturing} property).  HRSC schemes are also known for
giving stable and sharp discrete shock profiles and for having a high
order of accuracy, typically second order or more, in smooth regions
of the solution.

In our code we perform the time update of system (\ref{hydro_system})
from $t^n$ to $t^{n+1}$
according to the following conservative algorithm:
\begin{eqnarray}
    {\bf u}_{i,j}^{n+1} = {\bf u}_{i,j}^{n}
    & - & \frac{\Delta t}{\Delta r}
    (\widehat{{\bf f}}_{i+1/2,j}-\widehat{{\bf f}}_{i-1/2,j}) 
\nonumber \\
& - & \frac{\Delta t}{\Delta \theta}
    (\widehat{{\bf g}}_{i,j+1/2}-\widehat{{\bf g}}_{i,j-1/2}) 
\nonumber \\
& + & \Delta t \,\, {\bf s}_{i,j} \, ,
\end{eqnarray}
\noindent
improved with the use of consecutive sub-steps to gain accuracy in
time (see Shu \& Osher 1989).  Index $n$ represents the time level and
the time (space) discretization interval is indicated by $\Delta t$ ($\Delta
r, \Delta\theta$).  The ``hat" in the fluxes denotes the so-called numerical
fluxes which are computed by means of an
approximate Riemann solver according to a generic flux-formula
(supressing index $j$):
\begin{equation}
\widehat{{\bf f}}_{i+1/2} = \frac{1}{2}
({\bf f}({\bf u}^L) + {\bf f}({\bf u}^R) - Q).
\end{equation}

The flux-formula makes use of the complete characteristic information
of the system, eigenvalues (characteristic speeds) and right and
left eigenvectors through the {\it viscosity term} $Q$.  This
information is used to provide the appropriate amount of numerical
dissipation to obtain accurate representations of discontinuous
solutions without excessive smearing, avoiding, at the same time, the
growth of spurious numerical oscillations associated with the Gibbs
phenomenon.  Generic expressions for the characteristic speeds and
right eigenvectors for the general relativistic hydrodynamic equations
can be found in Font et al (1999). The left eigenvectors have been
obtained by Ib{\'a}{\~n}ez (1998).

Notice that the numerical flux is computed at the cell interfaces
(e.g., $i\pm1/2$) using information from the left and right sides. The
state variables, ${\bf u}$, must be accordingly computed
(reconstructed) in advance at both sides of a given interface out of
the cell-centered quantities.  The computation of these variables
determines the spatial order of the numerical algorithm and, in turn,
controls the local jumps at every interface. If these jumps are
monotonically reduced the scheme provides more accurate initial
guesses for the (either exact or approximate) solution of the local
Riemann problems. A wide variety of cell reconstruction procedures is
available in the literature. As this issue is particularly relevant
for the simulations we report below, we review the most standard
choices in the following section.

In addition, our code has the ability of using different approximate
Riemann solvers: the Roe solver (Roe 1981), widely employed in
classical fluid dynamics simulations, with arithmetically averaged
states, the HLLE solver (Harten, Lax \& van Leer 1983; Einfeldt 1988)
and the Marquina solver (Donat \& Marquina 1996), which has been
extended to relativistic hydrodynamics by Donat et al (1998). We have
performed a detailed comparison of the different solvers finding good
overall agreement among them. Hence, for the final computations
reported here we have employed Marquina's scheme.

Further specific issues of the hydrodynamical code are in order: 1)
the reconstructed (left and right) variables are the physical
(primitive) variables ${\bf w}$, except for the internal energy
density, $\varepsilon$, as we use a zero-temperature EOS. From these, 
the remaining (extrapolated) variables are computed algebraically.  
2) As we are considering adiabatic evolutions, we only solve
for the first four equations of system (\ref{hydro_system}). The internal
energy (proportional to the rest-mass density) is obtained algebraically
using the EOS. 3) Once all variables have been reconstructed, the set of 
local Riemann problems (as a result of the discretization on a numerical 
grid) is solved using Marquina's flux-formula.  4) The numerical fluxes are 
computed independently for each direction and the time update of the
state-vector ${\bf u}$ is done simultaneously using a method of lines
in combination with a second-order (in time) conservative Runge-Kutta
scheme, as derived by Shu \& Osher (1989).  5) After the update of the
conserved quantities, the primitive variables must be reevaluated. As
the relation between the two sets is not in closed algebraic form, the 
update of the primitive variables is done using an iterative Newton-Raphson
algorithm.

\subsection{Cell reconstruction procedures}
%==========================================

We turn now to describe the different ways we have considered to
compute the ${\bf u}^{L,R}$ states at every side of a cell interface.
This is commonly referred to as the cell-reconstruction process. In
the code we have three basic ways of performing such cell
reconstruction.

\subsubsection{MUSCL}
%--------------------

First, we can use a second-order TVD scheme (Harten 1984) of the MUSCL
type (Monotonic Upstream Schemes for Conservation Laws, van Leer
1979).  The total variation of a given quantity $\{{\bf
  u}_i^n\}_{i=-\infty}^{\infty}$ on a numerical grid is defined as:
\begin{equation}
TV({\bf u}^n)=\sum_{i=-\infty}^{\infty} |{\bf u}_{i+1}^n-{\bf u}_i^n|.
\end{equation}
A 2-level scheme is called TVD if
\begin{equation}
TV({\bf u}^{n+1}) \leq TV({\bf u}^n).
\end{equation}
\noindent
The TVD property guarantees the suppresion of spurious oscillations
near discontinuities.

The code uses slope-limiter methods to construct second-order TVD
schemes by means of monotonic piecewise linear reconstructions of the
cell-centered quantities. In this scheme, ${\bf u}^R_i$ and ${\bf
  u}^L_{i+1}$ are computed to second-order accuracy as follows:
\begin{eqnarray}
{\bf u}^R_i &=& {\bf u}_i + \sigma_i (x_{i+\frac{1}{2}}-x_i)
\\
{\bf u}^L_{i+1} &=& {\bf u}_{i+1} + \sigma_{i+1} (x_{i+\frac{1}{2}}-x_{i+1})
\end{eqnarray}
\noindent
($x$ denotes a generic spatial coordinate. In the code it can be either
$r$ or $\theta$ depending on the direction the integration takes place).
Our choice for the slope is the standard minmod slope which provides
the desired second-order accuracy for smooth solutions, while still
satisfying the TVD property:
\begin{equation}
\sigma_i = \mbox{minmod}\left(\frac{{\bf u}_i-{\bf u}_{i-1}}{\Delta x},
\frac{{\bf u}_{i+1}-{\bf u}_{i}}{\Delta x}\right),
\end{equation}
\noindent
where the minmod function of two arguments is defined by:
\[ 
\mbox{minmod}(a,b) = \left\{ \begin{array}{cl}
     a  & \mbox{if $|a|<|b|$ and $ab>0$} \\
     b  & \mbox{if $|b|<|a|$ and $ab>0$} \\
     0  & \mbox{if $ab\leq 0$}
                  \end{array}
                  \right. 
\]
\noindent
Notice that setting $\sigma_i=0$ provides a first-order piecewise constant
reconstruction scheme. This is then equivalent to the original scheme
by Godunov (1959).

\subsubsection{ENO}
%------------------

Our second choice involves the use of high-order (up to third-order) ENO
schemes (Harten \& Osher 1987; Harten et al. 1987). In these schemes, 
given ${\bf u}_i$, the cell average
of a piecewise smooth function, one must construct a piecewise
polynomial function of $x$ of uniform polynomial
degree $r-1$, $R(x;{\bf u})$, of the form:
\begin{equation}
R(x;{\bf u}_i)=\sum_{l=1}^{r-1} \frac{1}{l!}  b_{i,l} (x-x_i)^l
\end{equation}
\noindent
with $x\in[x_{i-1/2},x_{i+1/2}]$ and $b_{i,l}$ being the
corresponding polynomial coefficients. This reconstruction polynomial is
essentially non-oscillatory in the sense that it satisfies
\begin{equation}
TV(R(x;{\bf u}^{n+1})) \leq TV(R(x;{\bf u}^n)) + {\cal O}(\Delta x^r).
\end{equation}
\noindent
This property allows the ENO schemes to mantain the same order of
accuracy at local extrema (whereas the TVD schemes always drop
to first-order). The left and right states are then given by
\begin{eqnarray}
{\bf u}^R_i &=& R(x_{i+1/2};{\bf u}_i),
\\
{\bf u}^L_{i+1} &=& R(x_{i+1/2};{\bf u}_{i+1}). 
\end{eqnarray}
\noindent
The way the coefficients $b_{i,l}$ are computed for different order
reconstruction polynomials is explained in the appendix of Harten et al
(1987).

\subsubsection{PPM}
%------------------

Finally, we also use the third-order Piecewise Parabolic Method (PPM)
of Colella \& Woodward (1984). In its original design the PPM combined
the use of a parabolic reconstruction procedure with Godunov's exact
Riemann solver for (Newtonian) ideal gases. Here, we use the reconstruction
approach in combination with Marquina's flux formula.

In the PPM scheme the interpolated interface values ${\bf u}_{i+1/2}$ are
obtained using the quartic polynomial uniquely determined by the five zone
average values ${\bf u}_{i-2},\cdots,{\bf u}_{i+2}$ in the following way:

\begin{eqnarray}
&{\bf u}_{i+1/2}& = {\bf u}_i + \frac{\Delta x_i}{\Delta x_i + \Delta x_{i+1}}
({\bf u}_{i+1}-{\bf u}_i)  
\nonumber \\
&+&
\frac{1}{\sum_{j=-1}^{2}\Delta x_{i+j}} 
\nonumber \\
&\times&
\left\{ \frac{2\Delta x_{i+1} + \Delta x_i}
{\Delta x_i + \Delta x_{i+1}} \,\, {\bf X} \,\, ({\bf u}_{i+1}-{\bf u}_i) 
\right.
\nonumber \\
&-& 
\left.
\frac{\Delta x_{i-1} + \Delta x_{i}}{2\Delta x_i + \Delta x_{i+1}}
\,\, \Delta x_i \,\, \delta_m{\bf u}_{i+1} 
\right.
\nonumber \\
&+& 
\left.
\frac{\Delta x_{i+1} + \Delta x_{i+2}}{\Delta x_i + 2\Delta x_{i+1}}
\,\, \Delta x_{i+1} \,\, \delta_m{\bf u}_{i} \right\},
\end{eqnarray}
\noindent
with
\begin{eqnarray}
{\bf X} &=&
\frac{\Delta x_{i-1} + \Delta x_{i}}
{2\Delta x_i + \Delta x_{i+1}} -\frac{\Delta x_{i+2} + \Delta x_{i+1}}
{2\Delta x_{i+1} + \Delta x_{i}},
\end{eqnarray}
and
$\delta_m{\bf u}_{i} = \min(|\delta{\bf u}_i|,2|{\bf u}_i-{\bf u}_{i-1}|,
2|{\bf u}_{i+1}-{\bf u}_i|)
\times \,\, \mbox{sign}(\delta{\bf u}_i)$ if
$ ({\bf u}_{i+1}-{\bf u}_i)({\bf u_i}-{\bf u}_{i-1})>0$ and 0 otherwise.
Additionally
\begin{eqnarray}
\delta{\bf u}_i &=&
\frac{\Delta x_i}{\Delta x_{i-1}+\Delta x_i + \Delta x_{i+1}}
\nonumber \\
&\times& \left[\frac{2\Delta x_{i-1}+\Delta x_i}
{\Delta x_{i+1}+\Delta x_i}({\bf u}_{i+1}-{\bf u}_i)
\right.
\nonumber \\
&+&
\left.
\frac{\Delta x_{i}+2\Delta x_{i+1}}
{\Delta x_{i-1}+\Delta x_i}({\bf u}_{i}-{\bf u}_{i-1})\right].
\end{eqnarray}
\noindent
This algorithm provides a third-order accurate representation of 
${\bf u}_{i+1/2}$. For smooth flow solutions the left and right values
are then given by:
\begin{eqnarray}
{\bf u}^L_{i+1} = {\bf u}^R_i = {\bf u}_{i+\frac{1}{2}}.
\end{eqnarray}

In the presence of discontinuities these values are modified to
ensure the monotone character of the interpolation parabola. Colella and
Woodward (1984) proposed the following modifications:
${\bf u}^L_i={\bf u}^R_i={\bf u}_i$ if
$({\bf u}^r_i-{\bf u}_i)({\bf u}_i-{\bf u}^L_i)\leq 0$,
${\bf u}^L_i=3{\bf u}_i-2{\bf u}^R_i$ if 
$({\bf u}^R_i-{\bf u}^L_i)({\bf u}_i-0.5({\bf u}^L_i+{\bf u}^R_i))>
({\bf u}^R_i-{\bf u}^L_i)^2/6$ and
${\bf u}^R_i=3{\bf u}_i-2{\bf u}^L_i$ if 
$-({\bf u}^r_i-{\bf u}^L_i)({\bf u}_i-0.5({\bf u}^L_i+{\bf u}^R_i))>
({\bf u}^r_i-{\bf u}^L_i)^2/6$.

Additionally, the PPM scheme incorporates a special treatment of contact
discontinuities (contact steepening) as well as a ``flattening" procedure
to avoid spurious post shock oscillations. Details on these more
specialized aspects of the PPM scheme can be found in the original
reference of Colella \& Woodward (1984).

\subsection{Marquina's flux formula}
%===================================

The way the numerical flux at a given interface separating the states
${\bf u}^L$ and ${\bf u}^R$ is computed in Marquina's solver, is as
follows: We first compute the sided local characteristic variables and
fluxes
\begin{eqnarray*}
\begin{array}{rclcrcl}
\omega_{L}^p &=& {\bf{l}}^p({\bf{u}}_L) \cdot {\bf{u}}_L  &\quad&
\phi_{L}^p &=& {\bf{l}}^p({\bf{u}}_L) \cdot {\bf{f}}({\bf{u}}_L), \\
\omega_{R}^p &=& {\bf{l}}^p({\bf{u}}_R) \cdot {\bf{u}}_R  &\quad&
\phi_{R}^p &=& {\bf{l}}^p({\bf{u}}_R) \cdot {\bf{f}}({\bf{u}}_R),
\end{array}
\end{eqnarray*}
for $p=1,\ldots,5$.  Here ${\bf{l}}^p({\bf{u}}_L)$,
${\bf{l}}^p({\bf{u}}_R)$, are the (normalized) left eigenvectors of
the Jacobian matrices of the system.  Let
$\lambda_1({\bf{u}}_L),\ldots,\lambda_5({\bf{u}}_L)$ and
$\lambda_1({\bf{u}}_R),\ldots,\lambda_5({\bf{u}}_R)$ be their corresponding
eigenvalues. For $k=1,\ldots, 5$ the procedure is the following:

\begin{itemize}

\item
If    $ \displaystyle{\lambda_k({\bf{u}})}$ does not change sign in
  $[{\bf{u}}_L,{\bf{u}}_R]$, then the
 scalar scheme is upwind  \newline
    \newline
    \indent  \ \ \ \ \ \ \ \ \ \ \ \ \ If
    $\displaystyle{ \lambda_k({\bf{u}}_L)> 0}$ then \newline
    \indent  \ \ \ \ \ \ \ \ \ \ \ \ \ \ \ \ \ \ \
    $\displaystyle{\phi_+^k= \phi_L^k}$,  \newline
    \indent  \ \ \ \ \ \ \ \ \ \ \ \ \ \ \ \ \ \ \
    $\displaystyle{\phi_-^k= 0 }$, \newline
    \indent  \ \ \ \ \ \ \ \ \ \ \ \ \ else \newline
    \indent  \ \ \ \ \ \ \ \ \ \ \ \ \ \ \ \ \ \ \
    $\displaystyle{\phi_+^k= 0 }$, \newline
    \indent  \ \ \ \ \ \ \ \ \ \ \ \ \ \ \ \ \ \ \
    $\displaystyle{\phi_-^k= \phi_R^k}$,  \newline
    \indent  \ \ \ \ \ \ \ \ \ \ \ \ \ endif \newline
    \indent
\item
Otherwise, the scalar scheme is switched to the more viscous,
 entropy-satisfying, local-Lax-Friedrichs scheme \newline
 \newline
    \indent   \ \ \ \ \ \ \ \ \ \ \ \ \ \ $\displaystyle{
    \alpha_k= \max_{{\bf u} \in \Gamma
    ( {\bf u}_L ,{\bf u}_R)} |\lambda_k({\bf u})|}$, \newline
     \newline
     \indent   \ \ \ \ \ \ \ \ \ \ \ \ \ \
     $\displaystyle{\phi_+^k=.5(\phi_L^k+\alpha_k \omega_l^k)}$,
     \newline
     \indent    \ \ \ \  \ \ \ \ \ \ \ \ \ \
     $\displaystyle{\phi_-^k=.5(\phi_R^k-\alpha_k \omega_r^k)}$,
     \newline
     \indent
\end{itemize}
\noindent
$\Gamma( {\bf{u}}_L,{\bf{u}}_R)$ is a curve in phase space connecting
${\bf{u}}_L$ and ${\bf{u}}_R$.  In addition, $\alpha_k$ can be determined as
$$\alpha_k= \max\{|\lambda_k({\bf{u}}_L)|,|\lambda_k({\bf{u}}_R)|\}.$$

\noindent
Marquina's flux formula is then:
\begin{equation}
\label{mar}
\widehat{{\bf f}}_{i + 1/2} =
\sum_{p=1}^5 \left ( \phi^p_+ {\bf{r}}^p({\bf{u}}_L) +
          \phi^p_- {\bf{r}}^p({\bf{u}}_R) \right ),
\end{equation}
where ${\bf{r}}^p({\bf{u}}_L)$, ${\bf{r}}^p({\bf{u}}_R)$, are the
right (normalized) eigenvectors of the system. For further details on
this solver we refer the reader to Donat \& Marquina (1996).

\subsection{Numerical grid and boundary conditions}
%=======================================================

The code uses spherical polar coordinates $(r,\theta,\phi)$ and assumes
axisymmetry, i.e., there are no $\phi$ derivatives. The computational
domain in the radial direction extends from 0 to 1.2 times the radius
of the star.  In the polar direction the domain extends from 0 (pole)
to $\pi/2$ (equator). For the study of radial pulsations of spherical
stars we use fine grids of, typically, 400 radial grid-points and only 
one angular grid-point.  For
nonradial pulsations, for which the $\theta$-velocity does not vanish,
the typical grid-sizes we use are $80^2$ to $120^2$.
For studying quasi-radial pulsations in rotating stars, we
typically use more points in the radial direction than in the angular
direction.

We use a number of boundary grid-zones which depends on the stencil
size of the different schemes. Hence, for the MUSCL and ENO2 schemes,
we need to impose boundary conditions in two additional zones at each
end of the domain. For the ENO3 and PPM schemes, the number of
additional zones is four.  The boundary conditions are applied to
$\rho_0, v_r, v_{\theta}$ and $v_{\phi}$ and they are as follows: at the
center ($r=0$) the radial fluxes vanish. Hence, the radial velocity is
antisymmetric across the origin. The remaining variables are
symmetric. At the outer radial boundary, the ``atmosphere'' is re-set
to the initial data after each time step. At the pole ($\theta=0$) and
equator ($\theta=\pi/2$), all variables are symmetric, except for $v_\theta$,
since the initial data have equatorial-plane symmetry and, in this
paper, we assume that $v_{\theta}$ is a sum of ``polar'' (even) vector
harmonics.

%%%%%%%%%%%%%%%%%%%%%%%%%%%%%%%%%%%%%%%%%%%%%%%%%%%%%%%%%%%%%%%%%%%%%%%%%
%                                                                       %
\section{Code tests}                                                    %
\label{tests}                                                           %
%                                                                       %
%%%%%%%%%%%%%%%%%%%%%%%%%%%%%%%%%%%%%%%%%%%%%%%%%%%%%%%%%%%%%%%%%%%%%%%%%

\subsection{Shock tube test}
%===========================

We begin by testing the behaviour of the code and of its different schemes 
in a standard shock tube problem in flat spacetime. Although this is a
known test and the results are, to some extent, well documented in the
literature (see, e.g., Donat et al 1998, Mart\'{\i} and M{\"u}ller 1996) our
motivation to include them here is to show the correct implementation
of the different schemes in the present code, by comparing to a non
trivial exact solution with all types of nonlinear waves: a shock, a
contact discontinuity and a rarefaction wave.

\begin{figure}
\centering
\psfig{file=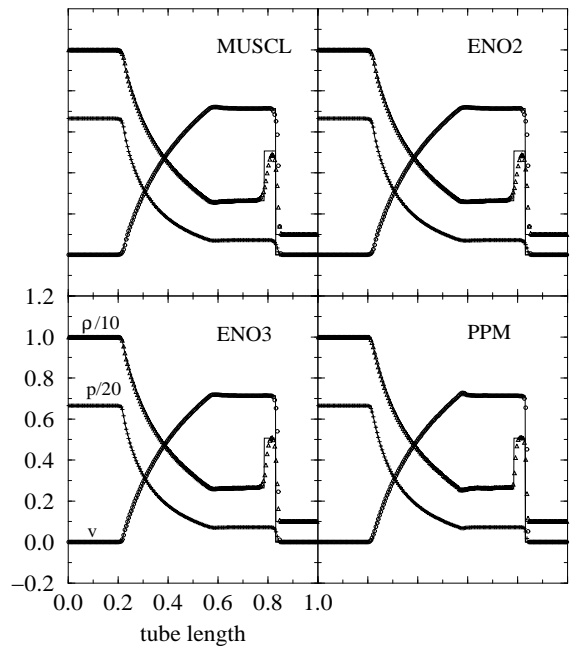,width=7.8cm,height=9.0cm}
\caption{
  Results for the shock tube test for the different implemented
  numerical schemes. The solid lines indicate the exact solution, at
  $t=0.4$. The symbols indicate the numerical solution for the
  velocity (circles), density (triangles) and pressure (plus signs).
  The density and pressure are scaled by a factor of 10 and 20
  respectively.}
\label{tube}
\end{figure}

Results for the shock tube test are plotted in Fig~\ref{tube}. The
initial discontinuous conditions correspond to problem 1 in Mart\'{\i}
and M{\"u}ller (1996).  The different panels show the final state, at
$t=0.4$, for the velocity, density and pressure for all numerical
schemes. The solid lines represent the exact solution and the symbols
indicate the numerical solution. We use a grid of 200 zones spanning a
domain of unit length.  The density and pressure are scaled by a
factor of 10 and 20 respectively. From this figure, we conclude that
the implementation of all schemes is correct as all of them reproduce
the exact solution and give the correct wave speeds.  The higher order
methods such as ENO3 and PPM give, as expected, the best results.

\subsection{Spherical stars}
%===========================

Next, we test our code in the non-rotating limit.  We study the
hydrodynamical evolution of non-rotating neutron star models in stable
equilibrium.  Since our axisymmetric code uses spherical polar
coordinates, it can also be run in essentially 1-D, by using only one
grid point in the angular direction. We determine radial pulsation
frequencies and compare to frequencies expected from perturbation
theory.  In 2-D, we study non-rotating stars that are perturbed from
equilibrium by an axisymmetric quadrupole perturbation and also
compare the pulsation frequencies to the ones obtained with a 
perturbative code.
 
We will focus our attention on two representative neutron star models:
Model 1 is a $N=1.5$ relativistic soft polytrope with $M/R=0.056$,
while model 2 is a more relativistic and stiffer polytrope with
$M/R=0.15$.  Table \ref{tab_models} lists the polytropic index $N$,
polytropic constant $K$, central density $\rho_c$, mass $M$ and
circumferential radius $R$ for the two models considered (all
quantities are dimensionless, using $c=G=M_{\odot}=1$).

\begin{table}
\begin{center}
\begin{tabular}{*{6}{r}}
\multicolumn{6}{c}{}\\
\multicolumn{6}{c}{\large \bf Initial models of spherical stars}\\ 
\multicolumn{6}{c}{}\\
\hline 
  & $N$ & $K$ & $\rho_c$ & $M$ & $R$ \\[0.5ex]
\hline
\\[0.5ex]
Model 1 & 1.5 & 4.349 & 8.10 $\times10^{-4}$ & 0.57 & 10.11 \\[0.5ex]  
Model 2 & 1.0 & 100   & 1.28 $\times10^{-3}$ & 1.40 & 9.59  \\[0.5ex]
\end{tabular}
\vspace{3mm}
\caption{
  Dimensionless ($c=G=M_\odot=1$) equilibrium properties of the two representative
  spherical neutron star models.}
\label{tab_models}
\end{center}
\end{table}

\subsubsection{1-D evolutions: radial pulsations}
%------------------------------------------------

The numerical evolution of initially static, non-rotating stars is
influenced by the size and sign of the truncation errors of the
hydrodynamical scheme. We observe the following characteristics, when
using HRSC schemes:

\begin{enumerate}
\item The truncation errors at the surface of the star initiate
  small-amplitude radial pulsations.
\item The radial pulsations are dominated by a set of discrete
  frequencies, which correspond to the normal modes of pulsation of
  the star.
\item The numerical viscosity of the finite-difference scheme damps
  the pulsations and the damping is stronger for the higher frequency
  modes.
\item The presence of a constant-density atmosphere affects the
  finite differencing at the surface grid-cells, which increases
  the numerical damping of pulsations and also can cause the
  star to drift to larger densities.
\end{enumerate}

Thus, we find that any hydrodynamical evolution of neutron star models,
with the present HRSC schemes, will be accompanied by small-amplitude
radial (or quasi-radial for rotating stars) pulsations.  The initial
amplitude of the radial pulsations and the small drift in density
converge to zero at the expected order rate, with increasing
resolution.  The value of the density in the atmosphere region can
have a large effect on the damping of the pulsations, if it is too
large. To minimize this effect, we typically set the density of the
atmosphere equal to $10^{-6}$ times the density of the central density
of the star.
 
\begin{figure}
\centering
\psfig{file=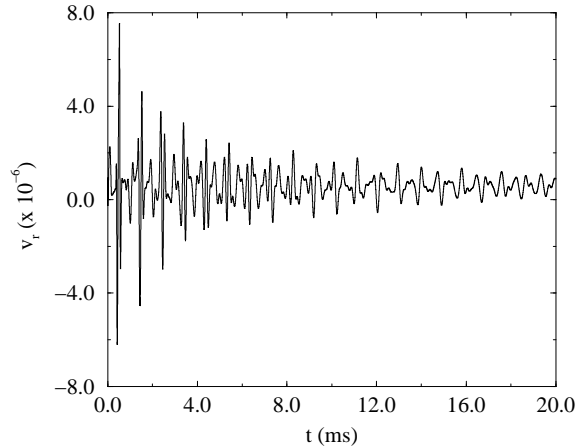,width=8cm}
\caption{
  Time evolution of the radial velocity of a spherical star (model 1). The
  radial pulsations are initiated by the truncation errors of the
  finite-difference scheme and are mainly a superposition of normal
  modes of the star.}
\label{fig_nonrot1}
\end{figure}

Fig. \ref{fig_nonrot1} shows the evolution of the radial velocity
$v_r$ at a radius of $0.25R$, with Model 1 as initial model. We use the
second-order MUSCL scheme with 400 radial grid-points.  The radial
velocity is initially a very complex function of time. As we will show
next, the pulsation consists mainly of a superposition of normal modes
of oscillation of the fluid.  The high frequency normal modes are
damped quickly and after 20ms the star pulsates mostly in its lowest
frequency modes. Because these oscillations are initiated by the
truncation errors, the magnitude of the radial velocity is extremely
small. The velocity oscillates around a non-zero residual, which 
also converges to zero as second order with increasing resolution.

\begin{figure}
\centering
\psfig{file=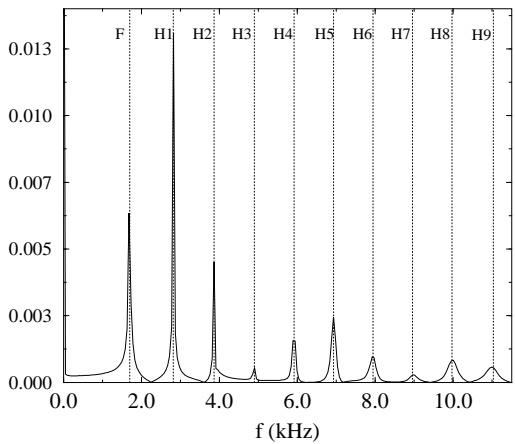,width=8cm}
\caption{Fourier transform of the evolution of the radial velocity in
  Fig. \ref{fig_nonrot1}. The frequencies are in excellent agreement
  with linear normal mode frequencies computed with an eigenvalue
  code (vertical dotted lines). The units of the vertical axis are 
  arbitrary.}
\label{fig_nonrot2}
\end{figure}

The small-amplitude radial pulsations in the nonlinear, fixed
spacetime evolutions correspond to linear normal modes of pulsation in
the relativistic Cowling approximation (McDermott, Van Horn \& Scoll 1983),
in which perturbations of the spacetime are ignored. A
Fourier transform of the density or radial velocity evolution can be
used to compute the normal mode frequencies.  Fig. \ref{fig_nonrot2}
shows the Fourier transform of the radial-velocity evolution shown in
Fig.  \ref{fig_nonrot1}. The normal mode frequencies stand out as
sharp peaks on a continuous background.  The width of the peaks
increases with frequency. The frequencies of radial pulsations
identified from Fig. \ref{fig_nonrot2} are shown in Table
\ref{tab_radial1}.
   
To compare the obtained frequencies to linear normal mode frequencies,
we use a code that solves the linearized relativistic
pulsation equations for the stellar fluid in the Cowling
approximation,
% (McDermott, Van Horn \& Scoll 1983; Lindblom \& Splinter 1990;
%Yoshida \& Eriguchi 1999) 
as an eigenvalue problem.  In Table \ref{tab_radial1}
we present the results of this comparison.  The typical agreement
between frequencies computed by the two methods is better than 0.5\%
for the fundamental $F$-mode and the lowest frequency harmonics
$H_1-H_4$ and better than 0.8\% for the the higher harmonics
$H_5-H_9$. We note that these frequencies are global, as they should
be for normal mode pulsations, i.e. the frequencies are the same at
any point in the star, in both the radial velocity and density
evolution.  This is a strong test for the accuracy of the evolution
code and our results can be used as a testbed computation for other
relativistic multi-dimensional evolution codes.

\begin{table}
\begin{center}
\begin{tabular}{*{4}{r}}
\multicolumn{4}{c}{}\\
\multicolumn{4}{c}{\large \bf Radial Pulsation Frequencies. Model 1}\\ 
\multicolumn{4}{c}{}\\
\hline 
Mode & nonlinear (kHz) & Cowling (kHz) & difference \\[0.5ex]
\hline
\\[0.5ex]
$F$  &    1.703  &        1.697   &    0.3\%   \\[0.5ex]
$H_1$ &    2.820  &        2.807   &    0.5\%   \\[0.5ex]
$H_2$ &    3.862  &        3.868   &    0.02\%  \\[0.5ex] 
$H_3$ &    4.900  &        4.910   &    0.2\%   \\[0.5ex]
$H_4$ &    5.917  &        5.944   &    0.4\%   \\[0.5ex]
$H_5$ &    6.930  &        6.973   &    0.6\%   \\[0.5ex]
$H_6$ &    7.947  &        8.001   &    0.7\%   \\[0.5ex]
$H_7$ &    8.960  &        9.029   &    0.8\%   \\[0.5ex]
$H_8$ &    9.973  &        10.057  &    0.8\%   \\[0.5ex]
$H_9$ &   11.030  &        11.086  &    0.5\%   \\[0.5ex]
\end{tabular}
\vspace{3mm}
\caption{Comparison of small-amplitude radial pulsation frequencies
  obtained with the present nonlinear evolution code to linear
  perturbation mode frequencies in the relativistic Cowling
  approximation. The equilibrium model is a nonrotating $N=1.5$
  relativistic polytrope with $M/R=0.056$.}
\label{tab_radial1}
\end{center}
\end{table}
 
\begin{figure}
\centering
\psfig{file=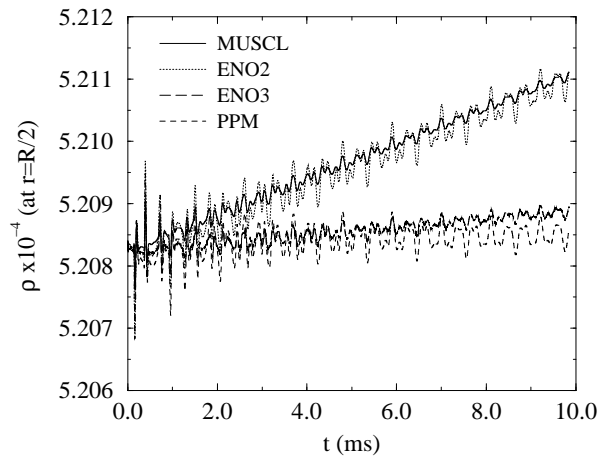,width=8cm}
\caption{Time evolution of the density at half-radius, initiated by  
  truncation errors, for a non-rotating initial model (model 2).  The
  small secular drift in density, when using second order schemes
  (MUSCL and ENO2), is significantly reduced or nearly eliminated,
  using third order schemes (ENO3 and PPM).}
\label{density_drift}
\end{figure}

While all four schemes give essentially the same radial pulsation
frequencies there are some differences in the hydrodynamical evolution
that are worth to be emphasized. Fig.~\ref{density_drift} shows the
density evolution, at $R/2$, for Model 2, using all four different
numerical schemes. The most striking difference is that the secular
drift in the density is much smaller for the third-order PPM and ENO3
schemes than for the second-order MUSCL and ENO2 schemes.  With a
resolution of 400 radial points, the drift is extremely small: after
$10ms$ the density has increased by only $0.3\%$. With the third-order
schemes, ENO3 and PPM, this drift is considerably smaller, being
practically unnoticeable in the case of PPM.  Another difference
among the schemes, is that each one of them excites the various normal
modes of pulsation with different amplitude in each mode. In ENO3, the
truncation errors at the surface also excite some very high-frequency
oscillations, in addition to low-order normal modes. These
high-frequency oscillations, however, do not represent a problem in
the identification of normal modes in a Fourier transform.  Finally,
the numerical damping rate of the excited pulsations is similar
in the four schemes, with ENO2 and PPM showing the smallest damping
rate.

\subsubsection{2-D evolutions: quadrupole pulsations}
%-----------------------------------------------------

Like radial pulsations, small-amplitude nonradial pulsations can be
studied with the present evolution code and the obtained frequencies
can be compared to perturbation results. We find that the truncation 
errors of the finite difference scheme do not excite nonradial 
oscillations with amplitudes large enough that one can identify them 
accurately in an evolution of unperturbed initial configurations. 
Instead, one has to perturb the static initial
configuration, using an appropriate eigenfunction for each nonradial
angular index $l$. For example, to excite the lowest-order quadrupolar
($l=2$) pulsations in a spherical star, we perturb the $v_\theta$ velocity
component, using an approximate eigenfunction with angular behaviour
same as the spherical harmonic function $Y_2^0$ and a simple $\sin(\pi
r/R)$ radial behaviour. The frequencies of the nonradial modes can
then be identified in a Fourier transform of the time-evolution of the
$v_\theta$ velocity component.
 
Table \ref{tab_quad2} shows a similar comparison, as in Table
\ref{tab_radial1}, for the quadrupole ($l=2$) pulsations of model 2.
Since the nonradial modes have to be computed on a two-dimensional
grid, we cannot use resolutions as high as in the 1-D computations.
For a grid-size of $120 \times 60$ zones, the difference between
frequencies computed by the two methods is $1.7 \%$ for the
fundamental $f$-mode and better than $0.4 \% $ for the $p$-modes
$p_1-p_4$.  For this grid-size, frequencies higher than the $p_4$
mode could not be computed accurately, because the grid is to coarse
to resolve their eigenfunctions (higher harmonic eigenfunctions have a
larger number of nodes in the radial direction).

%\begin{table}
%\begin{center}
%\begin{tabular}{*{4}{r}}
%\multicolumn{4}{c}{}\\
%\multicolumn{4}{c}{\large \bf Quadrupole Pulsation Frequencies. Model 1}\\
%\multicolumn{4}{c}{}\\
%\hline
%Mode & nonlinear (kHz) & Cowling (kHz) & difference \\[0.5ex]
%\hline
%\\[0.5ex]
%$f$   &    1.28   &  1.286  &    0.5\%         \\[0.5ex]
%$p_1$  &    2.68   &  2.681  &    0.04\%         \\[0.5ex]
%$p_2$  &    3.65   &  3.699  &    1.3\%         \\[0.5ex]
%$p_3$  &    4.66   &  4.719  &    1.3\%         \\[0.5ex]
%$p_4$  &    5.66   &  5.742  &    1.4\%            \\[0.5ex]
%$p_5$  &    6.83   &  6.764  &    1.0\%       \\[0.5ex]
%$p_6$  &    7.80   &  7.788  &    0.2\%       \\[0.5ex]
%\end{tabular}
%\vspace{3mm}
%\caption{Comparison of small-amplitude quadrupole ($l=2$) pulsation
%  frequencies obtained with the present nonlinear evolution code to
%  linear perturbation mode frequencies in the relativistic Cowling
%  approximation. The equilibrium model is a nonrotating $N=1.5$
%  relativistic polytrope with $M/R=0.056$.}
%\label{tab_quad1}
%\end{center}
%\end{table}

\begin{table}
\begin{center}
\begin{tabular}{*{4}{r}}
\multicolumn{4}{c}{}\\
\multicolumn{4}{c}{\large \bf Quadrupole Pulsation Frequencies. Model 2}\\
\multicolumn{4}{c}{}\\
\hline
Mode & nonlinear (kHz) & Cowling (kHz) & difference \\[0.5ex]
\hline
\\[0.5ex]
$f$   &  1.852     & 1.8843   &   1.7\%        \\[0.5ex]
$p_1$  &  4.095     &  4.1099   &  0.4\%        \\[0.5ex]
$p_2$  &  6.009     &  6.0351  &   0.4\%        \\[0.5ex]
$p_3$  &  7.858     &  7.8733  &   0.2\%        \\[0.5ex]
$p_4$  &  9.683     &  9.6740   &  0.1\%        \\[0.5ex]
\end{tabular}
\vspace{3mm}
\caption{
  Comparison of small-amplitude quadrupole ($l=2$) pulsation
  frequencies, obtained with the present nonlinear evolution code, to
  linear perturbation mode-frequencies in the relativistic Cowling
  approximation. The equilibrium model is a nonrotating $N=1.0$
  relativistic polytrope with $M/R=0.15$ (Model 2).}
\label{tab_quad2}
\end{center}
\end{table}

\begin{figure}
\centering
\psfig{file=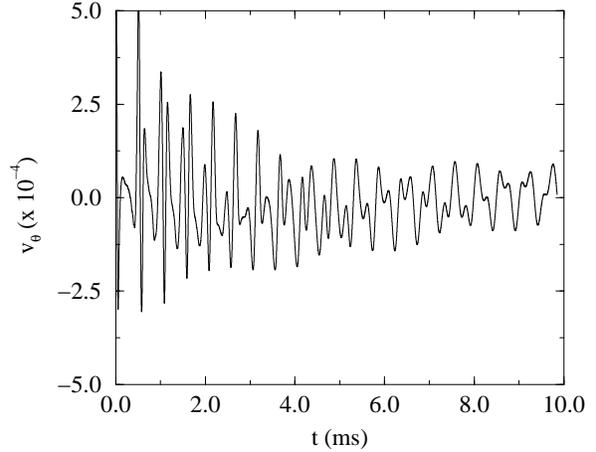,width=8cm}
\caption{
  Time evolution of the $v_\theta$ velocity component at $0.25R$, for a
  perturbed nonrotating initial model (Model 2) using the PPM scheme
  and an $(r,\theta)$-grid of $120\times 60$ points. The evolution is mainly a
  sum of the lowest-order quadrupolar pulsation modes of the star,
  excited using an approximate low-amplitude eigenfunction.}
\label{velz_sph2_nr}
\end{figure}

Fig.~\ref{velz_sph2_nr} shows the evolution of the $v_\theta$ at a radial
distance of $0.25R$, for model 2, perturbed with an approximate
quadrupolar egenfunction of small amplitude. In this evolution, the
PPM scheme was used with an $(r,\theta)$-grid of $120\times 60$ points.  The
evolution is mainly a sum of the lowest-order quadrupolar pulsation
modes of the star and allows for the accurate identification of
the $l=2$ normal-mode frequencies.

\subsection{Rotating stars}
%==========================

We now turn to the evolution of initially stationary, uniformly
rotating neutron stars. In the rotating case, we focus on a rapidly
rotating model, with angular velocity equal to $92\%$ of the
mass-shedding (Kepler) limit and equation of state and central density
parameters same as model 2 in Table \ref{tab_models}.  In these
evolutions, we observe the same qualitative properties as for
non-rotating stars and an additional important property: {\it the
rotation law (angular-momentum distribution), changes near the
stellar surface, due to the truncation errors of the finite
difference scheme}.  This can be
attributed to mainly two reasons: (i) the velocity component $v_\phi$ of
the fluid has a maximum at the surface and the second-order TVD
scheme, for example, although high-order accurate in smooth regions of
the solution, reduces to only first-order at extrema of the reconstructed
variables. ENO schemes, on the other hand, retain the order at local
extrema.  However, as we show below, preserving the angular momentum
distribution near the surface of the star, is still problematic with
such schemes.  This points to a second reason, for this behaviour.
(ii) The code evolves the relativistic momenta ($S_i$) and the
velocity components (as well as the other ``primitive" variables) must
be recovered through a root-finding procedure, which involves dividing
by the density (see Mart\'{\i} \& M{\"u}ller (1996) for details of this
procedure). At the surface of the star (where the density is very
small) this contributes to obtaining less than second-order accuracy.

\begin{figure}
\centering
\psfig{file=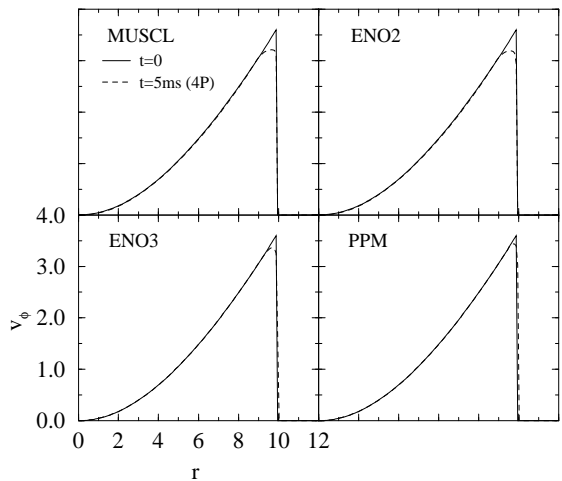,width=7.5cm,height=6.4cm}
\caption{Evolution of the velocity component $v_\phi$ of a rotating star.
The profiles show the radial dependence in the equatorial plane ($\theta=\pi/2$).
Dimensionless units are used in both axes.}
\label{f_rot1}
\end{figure}

\begin{figure}
\centering
\psfig{file=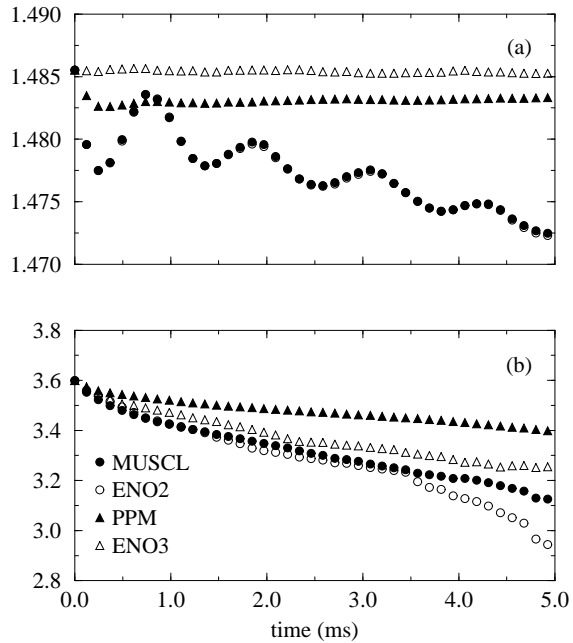,width=7.5cm}
\caption{Evolution of $v_\phi$ at two different radii of a rotating star, 
for the four different schemes: (a) $r=6$ (interior), (b) $r=9.8$ (surface).
The values are measured in the equatorial plane ($\theta=\pi/2$).
In the interior of the star the third-order schemes (ENO3 and PPM) retain
accurately the initial values while the second-order schemes (ENO2 and
MUSCL) show a very small secular drift ($0.8\%$ after $5ms$) accompanied by 
oscillations. This drift is due to finite-differencing truncation errors.
Both second-order schemes behave almost identically. Near the surface,
$r=9.8$ (bottom panel), the drift is more pronounced, being smallest for the 
third-order PPM scheme.}
\label{max_drop}
\end{figure}

A typical example of the evolution of a rotating star is presented in
Fig.~\ref{f_rot1}, which shows the evolution of the velocity component
$v_\phi$ obtained with the four schemes and with a grid-size of $160\times
40$ zones in $r$ and $\theta$ respectively. Depicted is the initial
equilibrium solution (solid line) as a function of the radial
coordinate (in the equatorial plane) and the final configuration,
after an evolution time of $5ms$, which corresponds to 4 rotational
periods.  The figure shows that $v_\phi$ remains close to its initial
value in the interior of the star.  But, near the surface, $v_\phi$
decreases with time. It is evident from this figure that there are significant
differences between second-order (top panels) and third-order (lower
panel) schemes. For the same resolution, the third-order schemes
result in a more accurate preservation of the initial rotation law,
both in the interior and the surface of the star. We note that, in the
interior, all schemes give satisfactory results, but for the
third-order schemes, the difference between the initial and final
solution is negligible. The more accurate preservation of the rotation
law for the third order schemes is emphasized in Fig.~\ref{max_drop},
which shows the change in $v_\phi$ at the interior ($r=6$, top panel)
and at the surface of the star ($r=9.8$, bottom panel), as a
function of time, for the four difference schemes.  
The values depicted correspond to the equatorial plane of the star 
($\theta=\pi/2$).
At the interior, the third-order schemes (ENO3 and PPM) retain accurately 
the initial rotation law (PPM shows a tiny initial jump to a slightly lowest 
value), while the second-order schemes (ENO2 and MUSCL) show a very small 
secular drift ($0.8\%$ after $5ms$) accompanied by oscillations. Correspondingly,
at the surface, bottom panel of Fig.~\ref{max_drop}, all schemes show a
nearly linear change in $v_{\phi}$, with the third-order PPM scheme
having the smallest slope.
 
%By comparing evolutions with different grid sizes, we find that
%the change of the angular momentum distribution improves as first-order 
%with resolution near the surface of the star and as second order
%with resolution in the interior.

\begin{figure}
\centering
\psfig{file=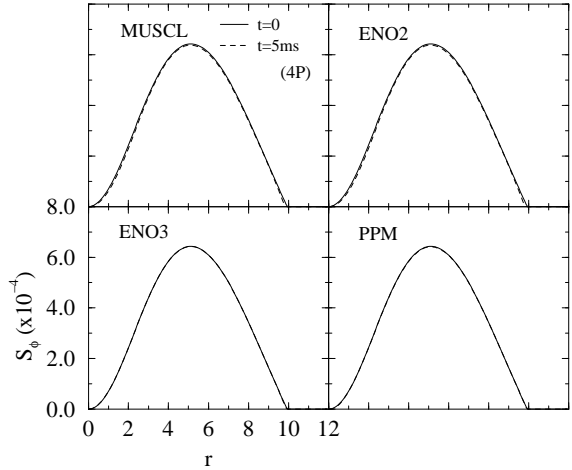,width=7.5cm,height=6.3cm}
\caption{Evolution of the momentum component $S_\phi$ of a rotating star.
The profiles show the radial dependence in the equatorial plane ($\theta=\pi/2$).
Dimensionless units are used in both axes.}
\label{f_rot2}
\end{figure}

In Fig.~\ref{f_rot2} we show the momentum component $S_\phi$ of the same
rotating star for the four numerical schemes as in Fig.~\ref{f_rot1}.
This is one of the conserved quantities {\bf u} directly evolved with
the code. As mentioned previously, the recovery of the primitive
variables is done via a root-finding procedure which involves dividing
by the density. Both quantities, $S_\phi$ and $\rho$, are very close to
zero around the surface layer and hence the procedure is very sensitive to
truncation errors.  Whereas the four schemes give accurate results for
the evolution of $S_\phi$, the computation of $v_\phi$ near the surface of
the star is only first-order accurate irrespective of the numerical
method.

\begin{figure}
\centering
\psfig{file=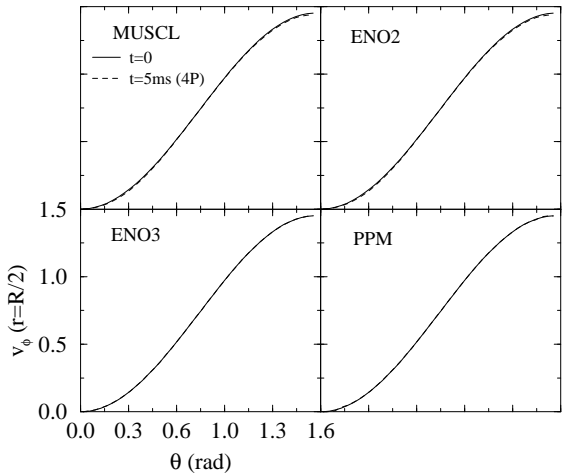,width=7.5cm,height=6.3cm}
\caption{Evolution of the velocity component $s_\phi$ of a rotating star. The
profiles show the angular dependence at half the radius of the star.}
\label{f_rot3}
\end{figure}

We plot in Fig.~\ref{f_rot3} the initial and final configurations of
the azimuthal component of the velocity as a function of the angular
coordinate, $\theta$, at half the radius of the star. The solid line
represents the initial solution and the dashed line corresponds,
again, to a final solution after $5ms$ (four rotational periods). All
schemes give excellent accurate results, most notably ENO3 and PPM.

\begin{figure}
\centering
\psfig{file=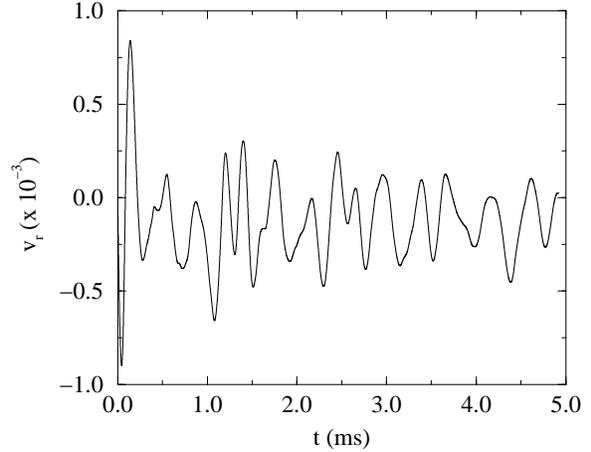,width=8cm}
\caption{Evolution of the radial velocity at $0.25R$
  for a rotating initial model (Model 2) using the PPM scheme
  and a $(r,\theta)$-grid of $160\times 40$ points. This figure shows
  the quasi-radial pulsations of the star as a result of a external
  perturbation.}
\label{velr_rot_qr}
\end{figure}

Quasi-radial pulsations of rotating relativistic stars have only been
studied in the slow-rotation limit (but without the assumption of a
fixed spacetime) by Hartle \& Friedman (1975) (see also Datta et al.
1998). We can study such pulsations for rapidly rotating neutron
stars, using appropriate radial eigenfunctions to excite them. In
Fig.~\ref{velr_rot_qr} we show the time evolution of the radial
component of the velocity (in the equatorial plane) at $0.25R$, after
an initial small-amplitude radial excitation of the stationary star
studied in this section. The quasi-radial pulsations can be followed
accurately for many dynamical times. 

We note that the computation of modes in a rotating star
requires an appropritate excitation, even for quasi-radial modes,
in order to obtain the corresponding frequencies with good accuracy in
a Fourier transform of hydrodynamical variables. A detailed study 
of quasi-radial modes in rapidly rotating relativistic stars of various
equations of state and rotation laws (in the Cowling approximation),
will appear elsewhere.

%\begin{table}
%\begin{center}
%\begin{tabular}{*{5}{r}}
%\multicolumn{5}{c}{}\\
%\multicolumn{5}{c}{\large \bf Quasi-radial Pulsation Frequencies. Model 2}\\
%\multicolumn{5}{c}{}\\
%\hline
%Mode & MUSCL & EON2 & ENO3 & PPM \\[0.5ex]
%\hline
%\\[0.5ex]
%$f$   &   2.45  &  & 2.45            \\[0.5ex]
%$h_1$  &   3.24  &  & 3.25           \\[0.5ex]
%$h_2$  &   4.28  &  & 4.27           \\[0.5ex]
%$h_3$  &       &    &   &        \\[0.5ex]
%$h_4$  &       &    &   &        \\[0.5ex]
%$h_5$  &       &    &   &        \\[0.5ex]
%$h_6$  &       &    &   &        \\[0.5ex]
%\end{tabular}
%\vspace{3mm}
%\caption{Frequencies (in kHz) of quasi-radial modes for the rotating relativistic
%  star of Model 2.}
%\label{qr_tab}
%\end{center}
%\end{table}

%%%%%%%%%%%%%%%%%%%%%%%%%%%%%%%%%%%%%%%%%%%%%%%%%%%%%%%%%%%%%%%%%%%%%%%%%
%                                                                       %
\section{Discussion}                                                    %
\label{discussion}                                                      %
%                                                                       %
%%%%%%%%%%%%%%%%%%%%%%%%%%%%%%%%%%%%%%%%%%%%%%%%%%%%%%%%%%%%%%%%%%%%%%%%%

We have developed a numerical code which integrates the equations of
general relativistic hydrodynamics to study pulsations of
rapidly-rotating relativistic stars in a fixed background spacetime.
The finite-difference code is based on a state-of-the-art approximate
Riemann solver (Donat \& Marquina 1996).  Our axisymmetric
relativistic hydrodynamical code is capable of accurately evolving rapidly 
rotating stars for many rotational periods.  We
find that, for non-rotating stars, small amplitude oscillations have
frequencies that agree with linear, radial and nonradial, normal mode 
frequencies in the Cowling approximation. This study has been performed 
using a representative set of second- and third-order TVD and ENO numerical 
schemes.

Modern HRSC numerical schemes (as the ones used in our code),
satisfying the ``total variation diminishing" property (Harten 1984),
are second-order accurate in smooth regions of the flow, but only
first-order accurate at local extrema. In our rotating star evolutions we
find that this results in a secular drift of the angular momentum distribution
near the surface of the star.  Essentially non-oscillatory schemes (Harten \&
Osher 1987), which retain the order at local extrema, were also used
to investigate the preservation of the initial rotation law near the surface,
with results similar to the second-order TVD methods.
The first-order accuracy obtained at the surface layer
irrespective of the method employed pointed out to the ill-posedness
of the primitive variables recovery procedure as the reason of the
angular momentum loss. The numerical scheme which has the smallest change
in the rotation law after many rotation periods is the third-order PPM
scheme.

It would be interesting to investigate if the computational cost of the
present code can be reduced with the use of surface-adapted coordinates or
fixed-mesh refinement and also to analyze whether the change of the
rotation law per rotation period will be significantly smaller in a 
frame co-rotating with the star.

The numerical findings reported in this paper are important for the study
of small-amplitude and nonlinear oscillations of rotating neutron stars such
as $f$, $p$ and $r$-modes oscillations.

Having adopted the PPM third-order scheme as our preferred choice for
studying hydrodynamical pulsations of rapidly-rotating stars, we plan to 
investigate axisymmetric pulsations of rotating proto-neutron stars, allowing
for various rotation laws and equations of state.

%%%%%%%%%%%%%%%%%%%%%%%%%%%%%%%%%%%%%%%%%%%%%%%%%%%%%%%%%%%%%%%%%%%%%%%%%
%                                                                       %
\section*{Acknowledgements}                                             % 
%                                                                       %
%%%%%%%%%%%%%%%%%%%%%%%%%%%%%%%%%%%%%%%%%%%%%%%%%%%%%%%%%%%%%%%%%%%%%%%%%

We thank Jos{\'e} M. Ib{\'a}{\~n}ez, Bernard F. Schutz and Ed
Seidel for helpful discussions, and John L. Friedman and Philippos
Papadopoulos for helpful comments on the manuscript.  
J.A.F acknowledges financial support
from a TMR grant from the European Union (contract nr.
ERBFMBICT971902). He also wants to thank Rosa Donat for useful
discussions on numerical schemes and for allowing us to use her ENO
cell-reconstruction routines.  K.D.K.  is grateful to the
Max-Planck-Institut f{\"u}r Gravitationsphysik (Albert-Einstein-Institut),
Golm, for generous hospitality.

\end{document}